# When Editors Revolt: Characterizing Journal Declarations of Independence

Saskia van Walsum[*], Lisa Matthias[**] , Juan Pablo Alperin[***] and Stefanie Haustein[****]

[*]*svanw056@uottawa.ca*

https://orcid.org/0009-0003-6331-1666

School of Information Studies, University of Ottawa, Canada

****lisa.matthias@hu-berin.de**

https://orcid.org/0000-0002-2612-2132

Berlin School of Library and Information Science, Humboldt-Universität zu Berlin, Germany

***juan@alperin.ca*

https://orcid.org/0000-0002-9344-7439

School of Publishing, Simon Fraser University, Canada

****stefanie.haustein@uottawa.ca*

https://orcid.org/0000-0003-0157-1430

School of Information Studies, University of Ottawa, Canada

**Abstract**

When editorial boards resign from their journals and publishers and declare their independence, two competing journals can result: the original journal under a new editorial board (a 'zombie' journal), and a new journal established by the departing editors (a 'breakaway'). The bibliometric community saw such an event when the board of *Journal of Informetrics* left Elsevier to found *Quantitative Science Studies*. We analyzed 39 breakaway-zombie journal pairs that have formed since 1989 and their declarations of independence to understand why and how they happen. Results show that declarations of independence were motivated by concerns related to governance and business model and overwhelmingly happened at journals owned by the Big Five publishers. Breakaway editors tended to found new journals at smaller publishers and adopt diamond publishing models. These findings suggest that dissatisfaction with commercial publishing models is growing, and that community-led alternatives can motivate change.

## 1. Introduction

Despite their flaws and limitations, academic journals are crucial scaffolding for our system of scholarly communication. They help define disciplines and set research agendas, underpin the system of checks and balances that protect scientific integrity, and are a means of disseminating, archiving, and evaluating research (Fyfe et al., 2017, 2017; Halliday, 2001; Lagoze et al., 2015).

Who controls journals, both from an academic and financial point of view, has meaningful consequences for how science is practiced and who is permitted to access and produce knowledge. Beginning in the 1980s, in what would become known as the “Serials Crisis”, publishing companies took advantage of inflexible demand for highly-regarded academic works to inflate subscription prices beyond what collections budgets could sustain (Douglas, 1990). So-called “Big Deals” promised to reduce the cost of subscriptions per journal, but also led to further consolidation of academic publishing markets and increased reliance on the large publishers that could offer them (Butler et al., 2023, Larivière et al., 2015**)**. The 2000s saw the growth of alternative responses from the academic community, from preprint servers to the Open Access movement advocating for innovation in publishing models (Fyfe et al., 2022). Throughout this sometimes-adversarial recent history of commercial publishers and academic communities, editorial boards have periodically taken matters into their own hands. Since 2015, RetractionWatch has reported at least fifty incidents of mass resignations by editorial boards (RetractionWatch, 2026). While each mass resignation did not necessarily result in a breakaway journal, the reasons for editorial rebellions often seem to echo broader concerns about the publishing industry. In the 1990s, for example, some editors rebelled against high subscription prices (for example, Rosenzewig, 2000). More recently, resignations have highlighted rising Article Processing Charges (APCs) and falling acceptance standards (for example, Keay, 2020; Dattaro, 2023). In some cases, declaring independence is the best option editors have to directly influence the direction of their journal (Sanderson, 2024)

When the editorial team of an existing journal collectively decides to leave their publisher and found a competing journal, the event is called a “declaration of independence” (Open Access Directory, 2024). The new journals have been referred to as "breakaway journals" (Wilson, 2016), while the original journals, now with a new editorial team, are known as "zombie journals" (Parker-Hay, 2023).

While breakaway journals periodically punctuate academic news and discussion boards, little is known about their long-term behaviour or their impact on academic publishing. As a commercial good, academic work stands out for its immutability - every journal, even every article, is a “tiny monopoly” in its field (Buranyi, 2017). A breakthrough research publication cannot be replicated more cheaply by a competitor. Therefore, as long as libraries are able to pay subscription costs, publishers are free to continue creating more journals or raising subscription prices (Buranyi, 2017; Larivière et al., 2015). Although in some ways breakaway journals could be said to be just like any other new, unindexed journal, they have a few distinct characteristics. Editors are the engines of an academic journal, setting its direction and deciding what work merits inclusion (Baccini & Re, 2025). Breakaway journals have the advantage of an established group of editors. Although it is true that they cannot compete to offer identical *content* as the zombie journal, if the journal’s prestige follows the editors and not the title, they may be able to compete for *standing* and *trust* from their academic community, and from there for readership. It has also been suggested that breakaway journals could have meaningful implications for the future of academic publication. They could be a way to flip journals from paywalled funding models to open ones (Laakso et al., 2016; Malakoff, 2000), or to increase competition in academic publishing markets (Buckholtz, 2001). Some of them have been vanguards of change, adopting diamond OA publishing models (Racy, 2025; Sanderson, 2023). There is some evidence that breakaways outperform their zombies in both longevity and impact, and that they can quickly become self-sustaining (Siler, 2026; Anderson, 2026; Blixrud, 2001; Buckholtz, 2001; Rosenzweig, 2000; Wilson, 2016). However, the evidence in support of journal declarations of independence as tools for breaking open academic publishing, or as significant challengers to the companies that dominate it, is largely anecdotal.

The same could be said of zombie journals. This is unfortunate, since their fates may have something to tell us not only about the relative effectiveness of their breakaways, but about the importance of declarations of independence for scientific communication overall. In one recent interview about the zombie journal *Neuroimage*, an Elsevier representative reported that the journal had since broadened its scope and seen a steady increase in publication output, as well as receiving increased submissions from authors in Asia and experiencing a slight dip in its CiteScore (Anderson, 2026). An analysis of 12 breakaway titles and their zombies suggested that the migration of established writers from the zombie to the breakaway opens room for high quality works from early-career researchers, leading to high impact factors in both journals (Carpenter, 2013). To the extent that this pattern holds for all zombie-breakaway pairs is unknown, but is it unlikely to be true across the board, given that, in some cases, editorial revolts are inspired by publisher behaviours that are reminiscent of a predatory journal, such concerns about degraded publication standards due to pressure to inflate publication volumes (for example, Kincaid, 2023; Weinberg, 2023). A later study of 3 declarations of independence in bibliometrics, lingusitics and mathematics identified a mild decline in citations for zombie journals immediately following a split alongside increased regional representation in the zombie journal, with a greater presence of China-based authors in particular (Siler, 2026). While anecdotes and a few case studies provide some sense of the dynamics at play when an editorial board declares independence, little is known about this phenomenon and its effect on the resulting journals or—critically—to the scholarly community that relies upon them.

This paper offers a preliminary collection and analysis of known breakaway journals, with the goal of providing a foundation for understanding these dynamics. In particular, we ask:

1. What are the primary motivating factors of journal declarations of independence?
2. Which publishers are most often abandoned? Which ones are most often joined?
3. Which publishing models do breakaway journals tend to leave and which do they tend to join?
4. Are there discipline-specific characteristics among breakaway journals? If so, what are they?

## 2. Data

The data was collected by building on the existing list of breakaway journals from Open Access Directory (OAD) (Open Access Directory, 2024). These journals, and the posts and articles reporting on them, were used to develop search strings in Factiva, Scopus, Web of Science, PubMed, Érudit, and Google. Searches were run in 2024 and again in 2025 to expand on the existing list through evidence including published announcements, news articles or academic society bulletins, and statements issued by the parties involved.

For the purposes of this search, a breakaway journal was considered to exist if:

1. Editors resign as a group from the original journal.
2. A new journal is established.
3. The composition of that new journal's board contains editors that resigned from the original.
4. An explicit link is made by the editors and/or their academic community between their resignation and the founding of a new journal.

Eight journals not meeting these criteria were removed from the original OAD list, while the search uncovered an additional eight breakaway journals. The final collection includes 39 instances of journal declarations of independence between the years 1989 and 2025.

## 3. Methods

For each of the 39 breakaway journals, the following information was taken from the documentation collected during the search:

- Year of the declaration of independence
- Zombie and breakaway journal titles
- Publishers of zombie and breakaway journals
- Zombie and breakaway publishing models
- Field of study
- Motive(s)

While variables like year of declaration, title, and publisher were considered factual information and therefore extracted by only one coder, publishing model and motive are more interpretive and were therefore double-coded. Details of coding strategy and coder agreement for discipline and motive are described in the corresponding sections below.

### *3.1 Publishing Models*

Publishing models were assigned to each journal based on information found in archived versions of journal and publisher websites from the year of declaration, using the Internet Archive's Wayback Machine. Journal publishing models were divided into the following categories, drawn from Open Access Network's description of funding models (2023):

- Closed: Readers pay to access the journal, submission is free
- Hybrid: Readers pay to access most of the journal, but authors may pay a fee to make their submission readable for free
- Gold: Authors pay an article processing fee for most articles, and all articles are available for free
- Diamond: Neither authors nor readers pay, and all articles are available for free; journal production costs are usually financed by a third party

*3.2 Fields of study*

Fields of study were assigned to one of the Broad and Second-level classifications of OECD Fields of Research and Development, as described in the 2015 Frascati Manual (OECD, 2015). Thirty journals found by the original 2024 search were double-coded for discipline, with 100% agreement between coders. Given the perfect agreement, the 9 additional journals collected during the subsequent round of searching in 2025 were assigned discipline categories by only one coder.

*3.3 Motives*

Each instance of a declaration of independence was tagged for at least one and at most two motives, as expressed by the breakaway editors. The categories of motive were:

1. **Editorial Standards**: Concerns related to the quality of a journal's content. This could include:
    - Unauthorized or unreviewed content
    - Poor or compromised peer review process
    - Scope or mission drift
    - Production errors affecting content integrity
    - Questionable special issues
2. **Governance:** Concerns related to who owns the journal or holds decision-making power. This could include:
    - Loss of editorial authority/ autonomy (e.g., publisher interfering with/ overriding decisions)
    - Legal or contractual conflict (e.g., disagreement about who owns the title)
    - Editorial appointments & board composition interference (e.g., hiring editors without approval, replacing Board members)
3. **Business Model**: Concerns related to how the journal prices its services or structures its business. This could include:
    - Pricing (including article processing charges)
    - Preference for a different funding model

- Profit allocation/ revenue concerns

4. **Workload:** Concerns related to the editors' workload. This could include:
   - Being asked to review too many submissions
   - An excess of administrative work

Two coders assigned motives to each of the 39 cases. Their agreement was calculated using Krippendorff's alpha, with Jaccard distance to account for the multiple possible values for each breakaway instance (Passoneau, 2006). Coders reached a Krippendorff alpha of 0.576, shy of the minimum threshold of 0.667 recommended by Krippendorf (Krippendorff, 2019). This may reflect the greater interpretive difficulty of identifying a secondary motive. When calculated on only the first assigned category, Krippendorff's alpha rises to 0.683, indicating that disagreement was concentrated in the selection of the secondary motive category rather than the primary one. Only three cases saw no overlap in motive codes. Coders were able to resolve all disagreements through discussion, with reference to the codebook.

## 4. Results

### *4.1 Behaviour over time*

Figure 1 shows the total number of declarations of independence per year, by the new publishing model of the breakaway journal. This data reflects the publishing model chosen when the breakaway journal was established; journals may have transitioned to a different model in later years. Breakaways appear to have happened in two clusters, with peaks in the early 2000s and again in the 2020s. From the late 1990s into the early 2000s, closed, i.e., paywalled, models dominated. But the surge of declarations in the late 2010s into the 2020s is accompanied by an apparent affinity for OA models, with diamond publishing being the dominant choice, followed by gold.

Figure 1: Breakaway journals, by new publishing model in the first year of operation

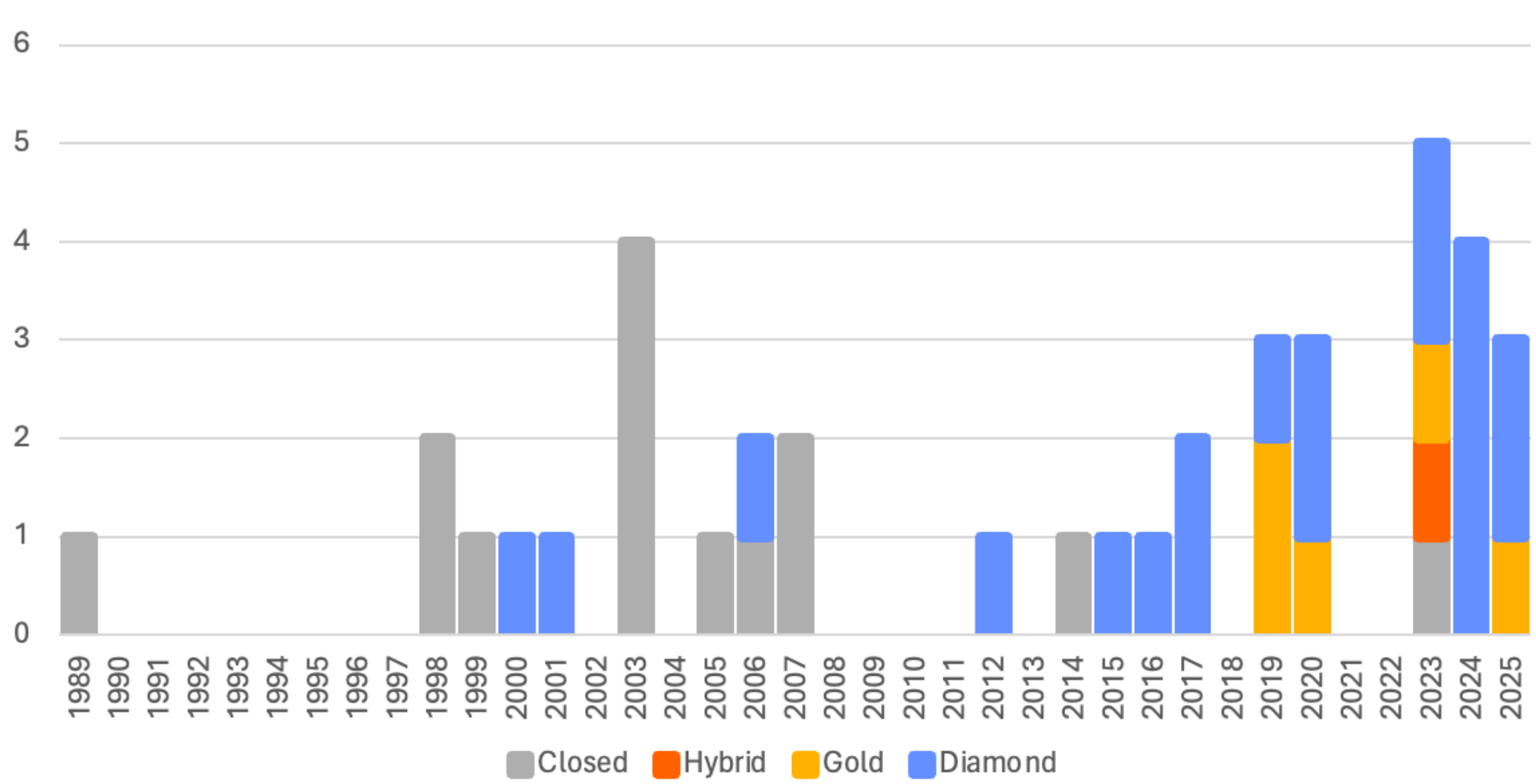

Business models have motivated declarations of independence in every year that saw any breakaway events, with governance and editorial standards receiving somewhat more attention in the 2020s. Of all breakaway journals, only one indicated editorial workload as a motivating factor. Figure 2 shows breakaway motives in five-year increments. Because up to two motives could be assigned to each breakaway, the number of motives in this figure is greater than the number of breakaway journals in that period.

Figure 2: Breakaway journal motives, in five-year increments

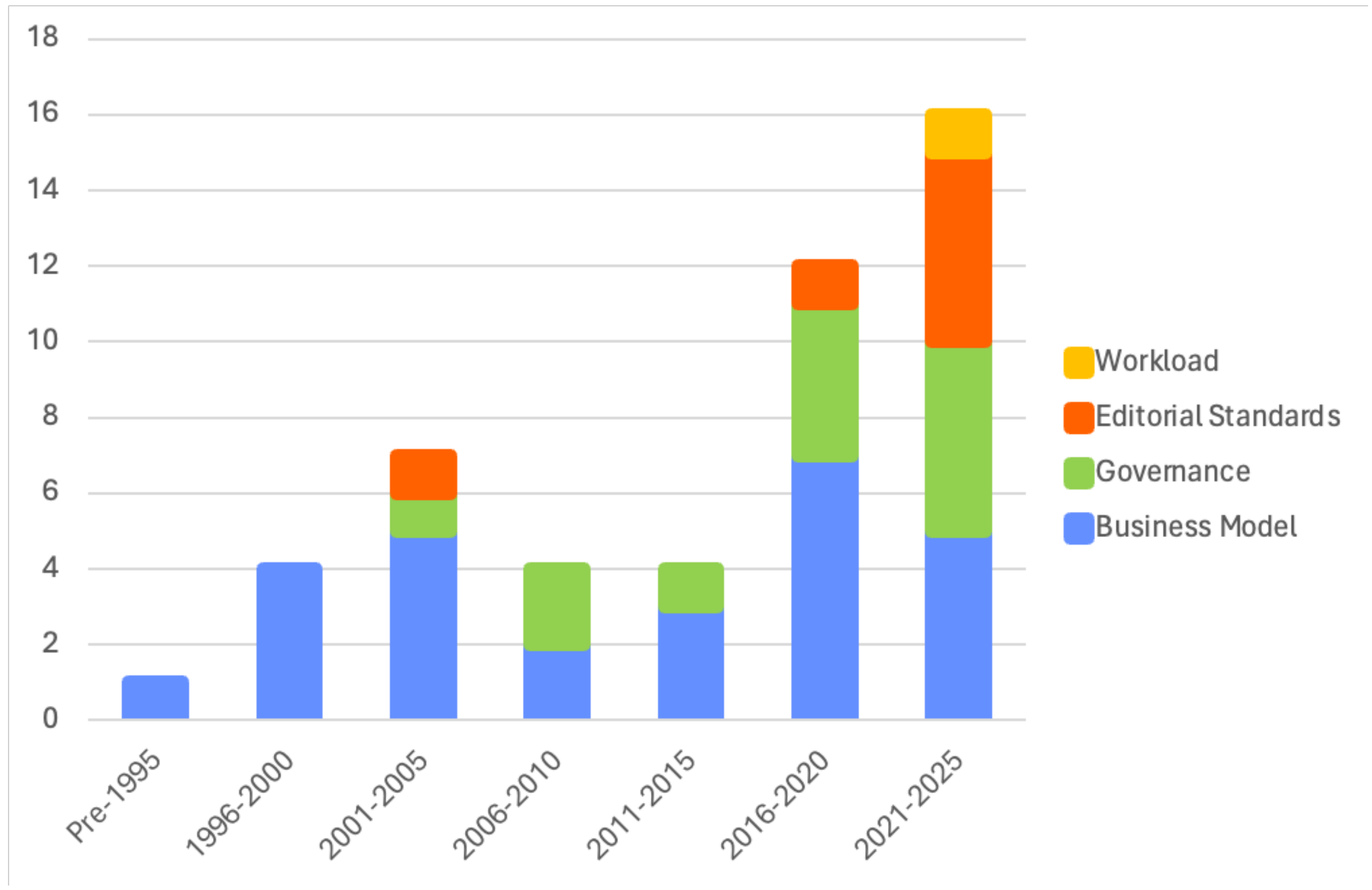

*4.2 Disciplines*

The natural sciences accounted for 17 breakaway journals (44%), followed by humanities and the arts with 10 instances (26%) and the social sciences with 8 (20%). Within the natural sciences, mathematics was the source of more than half of breakaways, with the same number of declarations of independence as all of the humanities and the arts. In all but 3 of the years to see a declaration of independence since the first in 1989 (van der Maarel, 1990), there has been at least one breakaway from the natural sciences. They were the main component of the cluster of declarations in the late 90s and early 2000s, while the recent spike combined natural science declarations with increased participation from humanities and the social sciences. Meanwhile, the agricultural and veterinary sciences fields (Broad Disciplines category 4 in the OECD classification) do not appear at all. As seen in Figure 3, motives appear broadly consistent across disciplines. Concerns about governance and about business models are apparent in almost every case. The social sciences are the only discipline category (besides the medical and health sciences, which only saw one declaration of independence) in which the governance motive was more prevalent than the business model motive.

Figure 3: Breakaways across disciplines, by motive

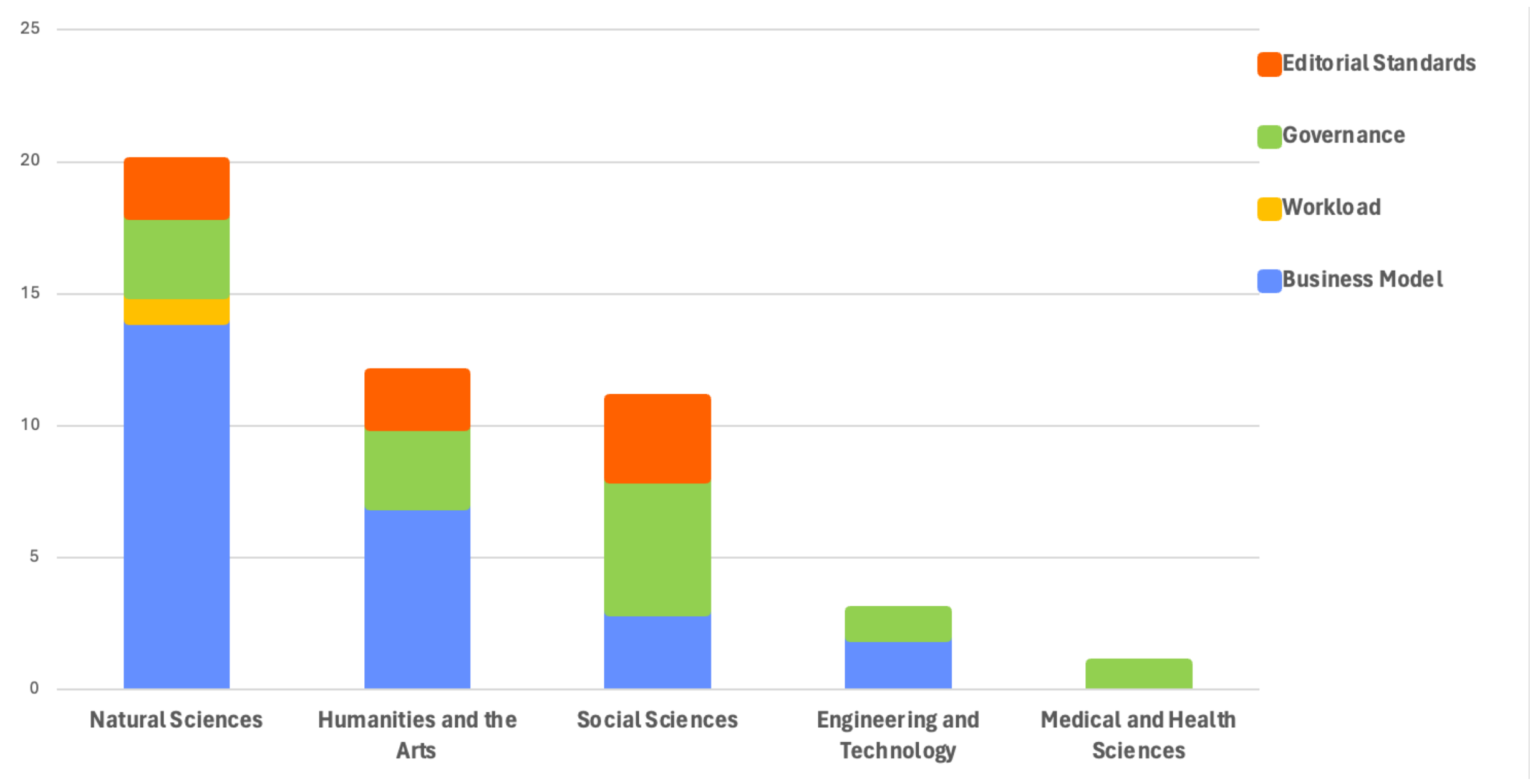

*4.3 Publication Models*

As noted above, business models are a significant concern for the editors of breakaway journals. Figure 4 compares the publishing model chosen by breakaway journals in their founding year to that of their zombie journal in the year of the editors' mass resignation. Editors are overwhelmingly choosing to declare independence from journals with closed and hybrid publishing models. While editors that leave closed journals tend to found breakaways that are also closed, declarations of independence from hybrid journals have primarily resulted in diamond breakaway publications.

Figure 4: Breakaway and zombie publishing models in the year of declaration

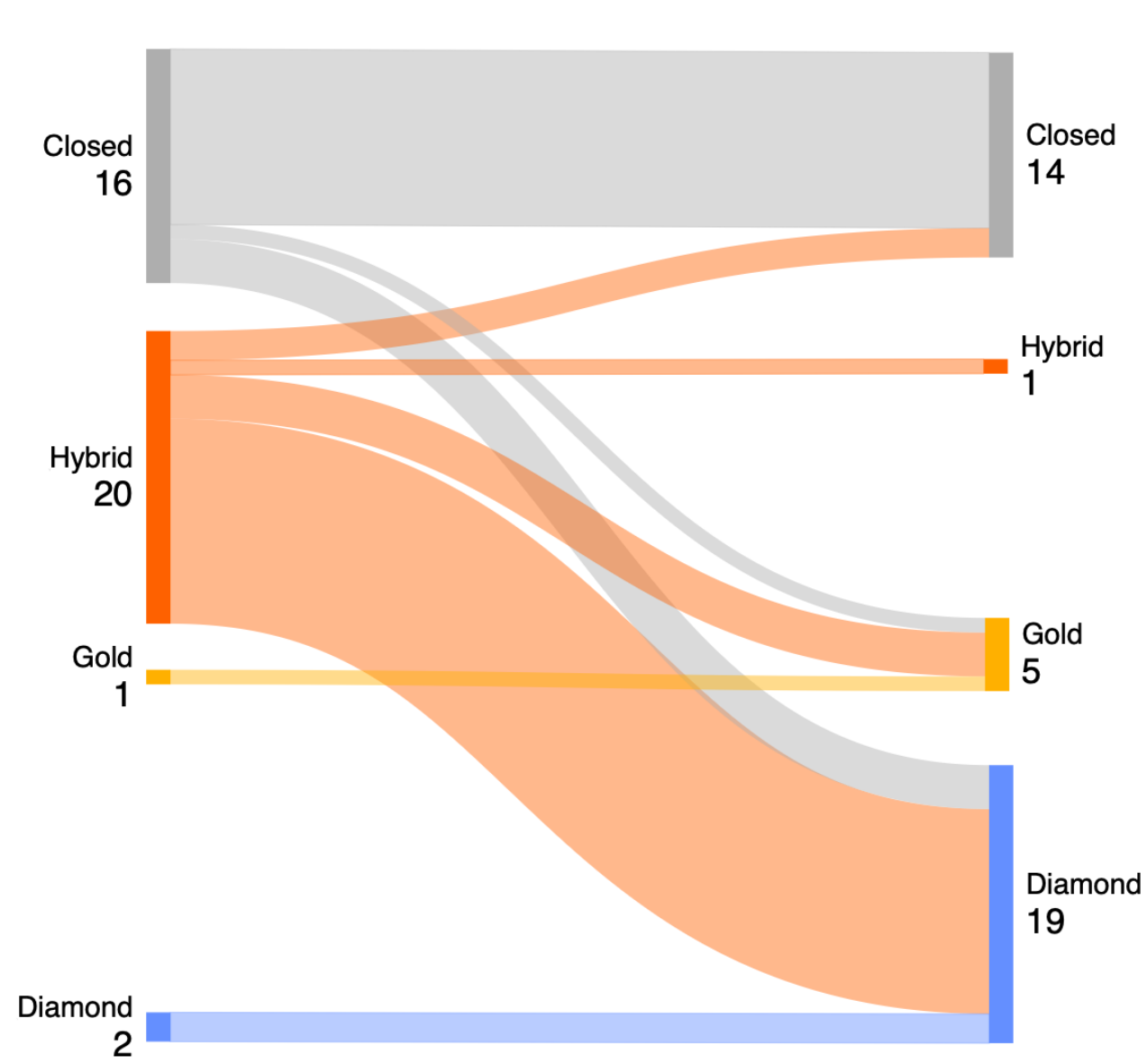

*4.4 Publishers*

The majority of journals in this study broke away from the quintet of large publishers known as the Big Five (Elsevier, Sage, Springer-Nature, Taylor & Francis and Wiley), and no editors declaring independence chose to join the Big Five. Figure 5 depicts the flow of journals from abandoned publishers to the destination publishers. The figure shows publishers as they were in the year of declaration, without accounting for later acquisitions on mergers. Breakaway journals that chose to affiliate with another organization were distributed across 9 university publishing platforms, 8 other publishers, and 2 associations. The majority of breakaway journals opted instead to affiliate with universities (either through libraries or university presses) or with other publishers. Among university-affiliated destination publishers, the most popular were Cambridge University Press and MIT Press, which welcomed 5 and 4 breakaways, respectively. The most popular alternative publisher destination is the Open Library of Humanities, which accepted 5 breakaways. Elsevier is by far the most-abandoned publisher, having lost 13 journals, a third of all breakaway journals. For a more complete depiction of flows by publisher, see Appendix 1.

Figure 5: Abandoned and destination publisher categories in the year of declaration

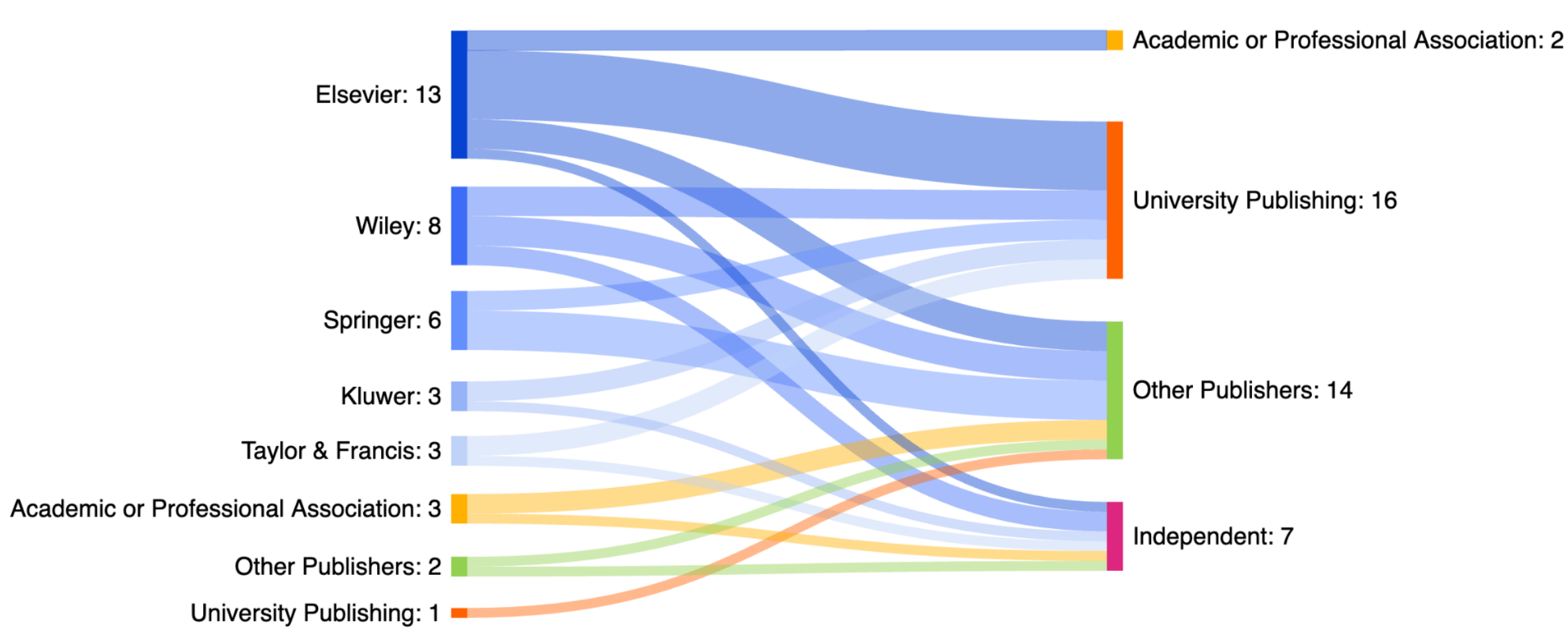

Although the small number of breakaway journals leaving non-Big Five publishers makes comparison difficult, concerns about business model account 58% of all motives identified in departures from the Big 5, while editorial standards weighed most heavily in decisions to leave university, association, or other publishers, accounting for 43% of all identified motives.

**5. Discussion**

We can see a strong tendency by editors who declare independence to move away from large commercial publishers and to adopt more open publishing models, especially diamond OA. The natural sciences, and mathematics in particular, have made up a large proportion of declarations of independence, with other disciplines joining the practice and contributing to spikes in the early 2000s and 2020s.

The motives of breakaway journals reflect the challenges to academic publishing at the time. In the 90s and 2000s, during the Serials Crisis, a majority of breakaway editors raised objections related to business models. In recent years, when the growing popularity of hybrid publishing has raised concerns about publishing integrity, business models remain a consistent issue, but are accompanied by concerns around governance and slipping editorial standards.

Editor's overwhelming tendency to leave the five largest publishers and their strong uptake of diamond publishing models suggests discontent with the dominant business models in commercial academic publishing. This is further supported by factors not reflected in this data.

Several early breakaway journals which remained closed (for example, Peet, 1999) pushed back against publishing norms of the time by slashing their subscription prices, sometimes to less than half of what was charged at the remaining zombie, or introduced contribution models not captured by the categories above, such as an optional program in which individual institutions could pay more for their subscription in order to fund access for less well-resourced users (van der Maarel, 2000).

Notable, too, is the fact that two of the most popular alternative publishers, MathOA and the Open Library of Humanities, have programs explicitly intended to support journals in their transition to diamond OA models (Math OA, 2024; Open Library of Humanities, 2025). Their relative success in the microcosmic world of breakaway journals speaks to the importance—and viability—of a community-driven transition to more open publishing.

When interpreting this data, it is important to keep in mind the many things it does not express. As mentioned above, the publishing model categories do not tell the full story of breakaway journals' potential to disrupt business models for academic journals. Publishing models and journal ownership may also change over time. An originally closed journal may since have transitioned to a more open model, or a diamond publication may have moved to a more closed structure. Some publishers that were independent at the time they were joined by a breakaway journal may since have been absorbed into the Big Five.

## 6. Open Science Practices

This paper builds on the work done and made openly available by Peter Suber, the OAD, OSTP, and Retraction Watch. In the spirit of continued openness, the dataset of journal titles, dates, publishers, publishing models, disciplines, and motives compiled for this research will be formatted for deposit and made openly available in the University of Ottawa Dataverse. Some documents which confirmed an instance of a breakaway journal are published in the pay-to-access archives of news publications and magazines. Therefore, the final dataset will not include copies of the supporting documents, but will list sources with links where available.

## 7. Acknowledgements

We thank Madelaine Hare for her help with first-round dataset coding.

## 8. Author Contributions

S.v.W.: Conceptualization, Data curation, Investigation, Methodology, Validation, Visualization, Writing—original draft

L.M.: Data curation, Investigation, Methodology

J-P.A.: Conceptualization, Funding acquisition, Methodology, Supervision, Writing—review & editing

S.H.: Conceptualization, Funding acquisition, Methodology,  Supervision, Writing—review & editing

## 9. Funding Information

This work was funded by the Volkswagen Foundation [Az.: 9C784]

**Appendix 1:** Abandoned and destination publishers in the year of declaration

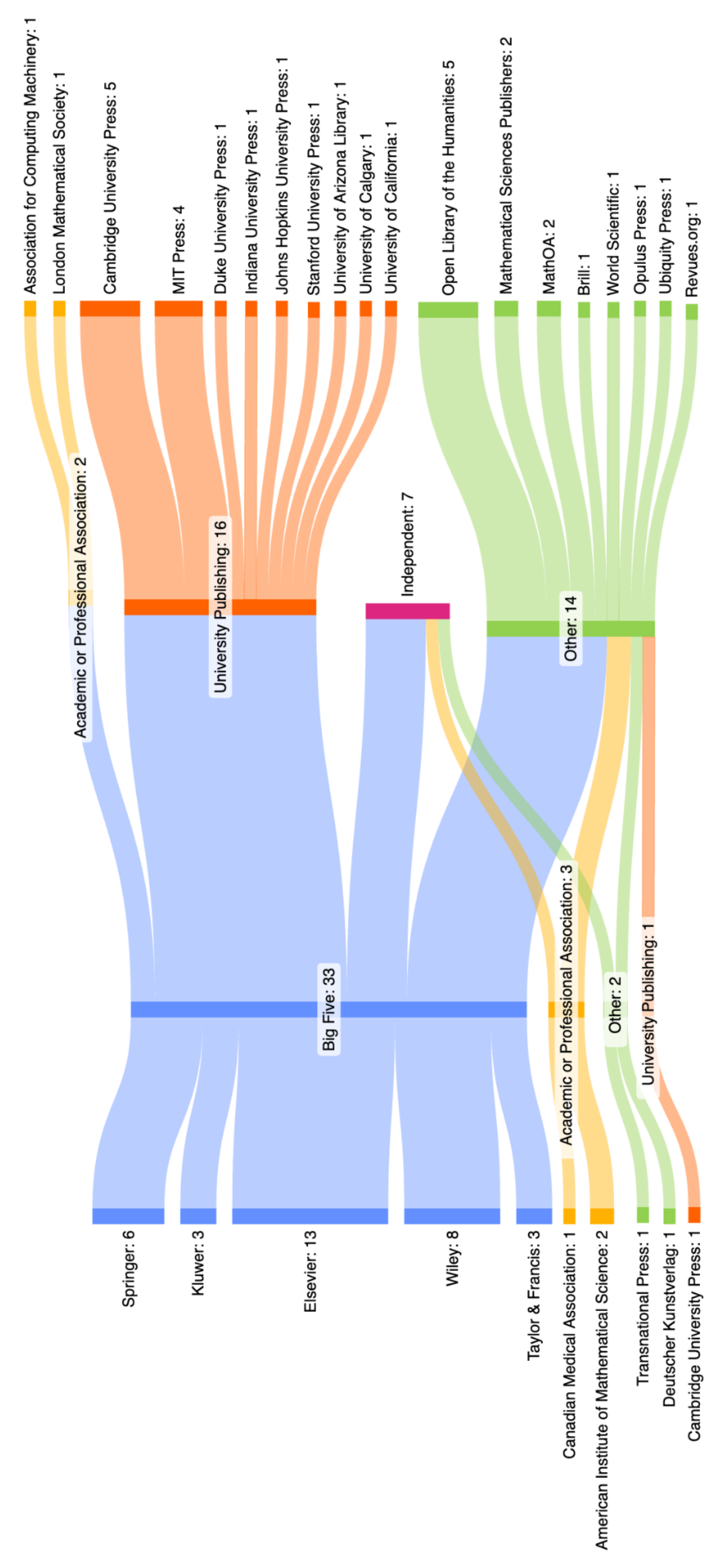